\documentclass[conference]{IEEEtran}
\ifCLASSINFOpdf
  \usepackage[pdftex]{graphicx}
\else
 \usepackage[dvips]{graphicx}
\fi
\graphicspath{{figures/}}
\usepackage[cmex10]{amsmath}
\DeclareMathOperator{\diag}{diag}
\usepackage{amsthm}
\usepackage{algorithmic}
\usepackage{array}
\usepackage{eqparbox}
\usepackage{txfonts}
\usepackage{verbatim, amssymb}
\usepackage{mathrsfs}
\usepackage{bm}
\usepackage{lipsum}
\usepackage{cite}
\usepackage{subfigure}
\usepackage{multirow}
\usepackage{epstopdf}

\usepackage[mathcal]{euscript}
\usepackage{amsmath}
\usepackage{amsthm}
\usepackage{amssymb}
\usepackage{graphicx}
\usepackage{algorithmic}
\usepackage{algorithm}
\usepackage{epsfig}
\usepackage{epstopdf}
\usepackage{booktabs}
\usepackage{colortbl}
\usepackage{amsmath}
\usepackage{xcolor}
\usepackage{graphicx}
\usepackage{cite} 
\newcommand{\myfigureshrinker}{\vspace{-0.1in}}

\providecommand{\mbf}[1]{\mathbf{#1}}						 
\providecommand{\bsym}[1]{\boldsymbol{#1}}				 

\colorlet{tableheadcolor}{gray!25} 
\newcommand{\headcol}{\rowcolor{tableheadcolor}} %
\colorlet{tablerowcolor}{gray!10} 
\newcommand{\rowcol}{\rowcolor{tablerowcolor}} %
\newcommand{\topline}{\arrayrulecolor{black}\specialrule{0.1em}{\abovetopsep}{0pt}%
            \arrayrulecolor{tableheadcolor}\specialrule{\belowrulesep}{0pt}{0pt}%
            \arrayrulecolor{black}}
\newcommand{\midline}{\arrayrulecolor{tableheadcolor}\specialrule{\aboverulesep}{0pt}{0pt}%
            \arrayrulecolor{black}\specialrule{\lightrulewidth}{0pt}{0pt}%
            \arrayrulecolor{white}\specialrule{\belowrulesep}{0pt}{0pt}%
            \arrayrulecolor{black}}

\DeclareMathOperator*{\argmax}{arg \, max}
\DeclareMathOperator*{\argmin}{arg \, min}

\begin{document}


\title{Impact of Radar and Communication Coexistence on Radar's Detectable Target Parameters}
\author{Haya Shajaiah, Ahmed Abdelhadi, and Charles Clancy\\
Bradley Department of Electrical and Computer Engineering\\
Hume Center, Virginia Tech, Arlington, VA, 22203, USA\\
\{hayajs, aabdelhadi, tcc\}@vt.edu\\
}
\maketitle
\footnotetext[1]{{\bf This work was sponsored by the Defense Advanced
Research Projects Agency (DARPA) under the Shared Spectrum Access for
Radar and Communications (SSPARC) program Contract DARPA-BAA-13-24.
The views expressed are those of the authors and do not reflect the
official policy or position of the Department of Defense or the U.S.
Government.\\
Distribution Statement "A" (Approved for Public Release, Distribution Unlimited)}}

\begin{abstract}
In this paper, we present our spectrum sharing algorithm between a multi-input multi-output (MIMO) radar and Long Term Evolution (LTE) cellular system with multiple base stations (BS)s. We analyze the performance of MIMO radars in detecting the angle of arrival, propagation delay and Doppler angular frequency by projecting orthogonal waveforms onto the null-space of interference channel matrix. We compare and analyze the radar's detectable target parameters in the case of the original radar waveform and the case of null-projected radar waveform. Our proposed spectrum-sharing algorithm causes minimum loss in radar performance by selecting the best interference channel that does not cause interference to the $i^{th}$ LTE base station due to the radar signal. We show through our analytical and simulation results that the loss in the radar performance in detecting the target parameters is minimal when our proposed spectrum sharing algorithm is used to select the best channel onto which radar signals are projected.
\end{abstract}
\IEEEpeerreviewmaketitle


\section{Introduction}
In recent years, bandwidth demands by commercial wireless operators increased rapidly. Smart phone users are now running large number of applications that require more networks resources. The limited available bandwidth for commercial wireless communications is becoming a challenge that will highly limit the growth of wireless devices in the future. A dynamic spectrum sharing among radio systems can improve data transmission capacity by acquiring the unused spectra \cite{Spectrum_Sensing}. Radar operations do not use frequency bands continuously in time and space. This provides good opportunity for spectrum sharing between radar and communication systems.

Federal agencies are now willing to share their spectrum with commercial users due to the high demand for spectrum by commercial operators. The 3550-3650 MHz band, currently used for military radar operations, is identified for spectrum sharing between military radars and communication systems, according to the NTIA's 2010 Fast Track Report \cite{NTIA10}. This band is very favorable for commercial cellular systems such as LTE Advanced systems. However, radar interference to cellular systems is a cause of concern for commercial operators and thus innovative methods are required to make spectrum sharing between radars and cellular systems a reality.

In the past, wireless systems were able to share government bands by operating on a low power to prevent the interference with the incumbent systems such as wireless local area network (WLAN) in the 5.25-5.35 and 5.47-5.725 GHz radar bands \cite{FCC_5GHz_Radar06}. Small cells operating in a low power have been proposed recently to operate in the 3.5 GHz radar band \cite{FCC12_SmallCells}.

To mitigate radar interference to LTE Advanced systems, a spatial approach for spectrum sharing between a MIMO radar and LTE cellular system with $N_{\text{BS}}$ base stations was proposed in \cite{Awais_Spatial}. Radar signals are manipulated such that they are not a source of interference to the LTE Advanced BSs. Because there exist many interference channels between the two systems, the interference channel with the maximum null space dimension is chosen based on the algorithm proposed by the authors, the radar signal is then projected onto the null space of that interference channel to mitigate interference to the LTE Advanced BS. This spatial approach results in small degradation in the radar performance \cite{SKC+12}.

In this paper, we consider a MIMO radar sharing spectrum with LTE cellular system that has $N_{\text{BS}}$ base stations. In order to mitigate radar interference, a spectrum sharing algorithm is proposed. The algorithm selects the best interference channel for radar's signal projection to mitigate radar interference to the $i^{\text{th}}$ BS. We consider a MIMO colocated radar mounted on a ship. Colocated radars have improved spatial resolution over widely-spaced radars \cite{MIMO_Radar}. The LTE cellular system operates in its regular licensed band and shares the $3.5$ GHz band with a MIMO radar in order to increase its capacity such that the two systems do not cause interference to each other. We focus on analyzing the performance of MIMO radars in detecting the radar's target parameters. We compare through analytical and simulation results the radar's detectable angle of arrival, propagation delay and Doppler angular frequency in two cases. In the first case, we estimate the radar's target detectable parameters using the original radar waveforms whereas we use null-projected radar waveforms for the estimation in the second case.
\vspace{-0.08in}
\subsection{Related Work}\label{sec:related}
In \cite{SKC+12}, the authors proposed a technique to project radar waveforms onto the null space of an interference channel matrix between the radar and the communication system. In their proposed approach, the cognitive radar is assumed to have full knowledge of the interference channel and modifies its signal vectors in a way such that they are in the null space of the channel matrix. In order to avoid interference to the communication system, a radar signal projection onto the null space of interference channel between radar and communication systems is presented in \cite{Beampattern}. In \cite{DH13}, a novel signal processing approach is developed for coherent MIMO radar to minimize the arbitrary interferences generated by wireless systems from any direction while operating at the same frequency using cognitive radio technology.

A resource allocation optimization problem with carrier aggregation is presented in \cite{Haya_Utility3} to allocate resources from the LTE Advanced carrier and the MIMO radar carrier to each user equipment (UE) in an LTE Advanced cell based on the running application of the UE. In \cite{Haya_Utility2} and \cite{Haya_Utility4}, the authors presented a resource allocation with users discrimination algorithms to allocate the eNodeB resources optimally among users and their applications. A resource allocation algorithm is proposed in \cite{Haya_Utility1} to allocate a primary and a secondary carriers resources optimally among users running real-time or delay-tolerant applications.
\subsection{Our Contributions}\label{sec:contributions}
Our contributions in this paper are summarized as:
\begin{itemize}
\item We present a spectrum sharing scenario between a MIMO radar and LTE system with multiple base stations and propose a channel-selection algorithm to select the best channel for radar's signal projection that maintains a minimum degradation in the radar performance while causing no interference to the LTE BS. We also present our null-space projection (NSP) algorithm that performs the null space computation.
\item We compare the radar's performance in estimating the target detectable parameters in the case of the original radar waveforms and the case of null-projected radar waveforms.
\end{itemize}

The remainder of this paper is organized as follows. Section \ref{section:arch} discusses the spectrum sharing scenario between MIMO radar and LTE cellular system. In Section \ref{sec:radar_main}, we describe colocated MIMO radars and then present the maximum likelihood (ML) estimate used as our performance metric for the MIMO radar system. In section \ref{sec:Algorithm}, we present our channel-selection and NSP algorithms and explain the projection of radar signal onto the null space of the selected interference channel. Section \ref{sec:Performance} provides mathematical analysis to study the effect of the projected radar signals on the estimate of the radar's target detectable parameters. In section \ref{section:sim}, we discuss simulation setup and provide quantitative results along with discussion. Section \ref{section:conclude} concludes the paper.
\section{System Model}
\label{section:arch}
\begin{figure}
\centering
\includegraphics[height=2in, width=3.4in]{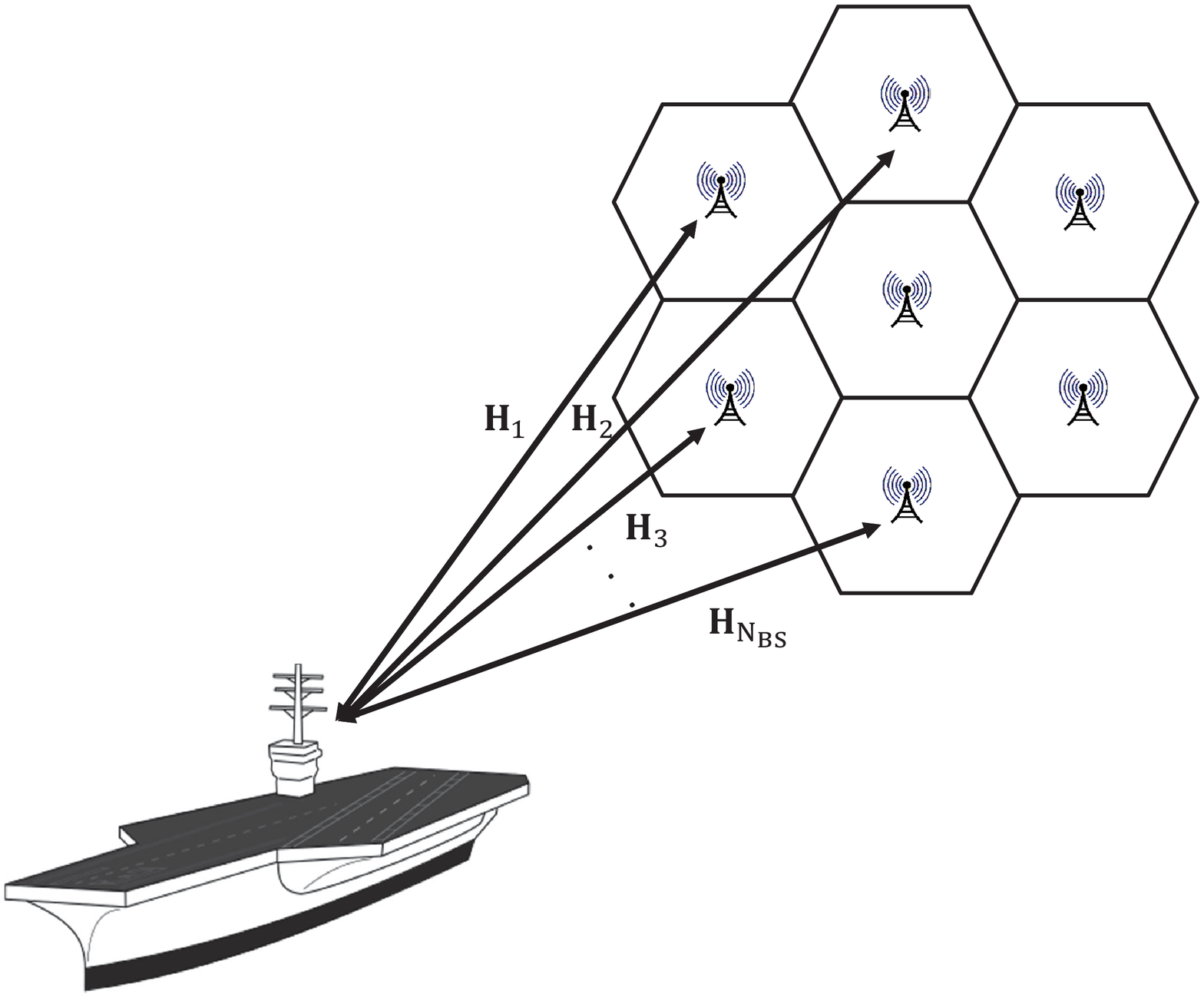}
\caption{Spectrum-sharing scenario between LTE cellular system and a maritime MIMO radar.}
\label{fig:System_Model_Arch}
\end{figure}

In this paper, we consider a colocated MIMO radar and a MIMO LTE communication system. The two systems are the primary users of the 3550-3650 MHz band under consideration. The MIMO radar has $M_T$ transmit antennas and $M_R$ receive antennas. The LTE communication system has $N_{\text{BS}}$ base stations, each BS is equipped with $N_T^{\text{BS}}$ transmit antennas and $N_R^{\text{BS}}$ receive antennas, with the $i^{\text{th}}$ BS supporting $K_i^{\text{UE}}$ user equipments (UE)s. Each UE is equipped with  $N_T^{\text{UE}}$ transmit antennas and $N_R^{\text{UE}}$ receive antennas. The colocated radars give better target parameter identifiability and improved spatial resolution as their antenna spacing is on the order of half the wavelength of the carrier \cite{MIMO_Radar}.
The MIMO radar projects its signal onto the null space of the interference channel while illuminating a target. The MIMO radar is sharing $N_{\text{BS}}$ interference channels $\mbf H_i^{N_R^{\text{BS}} \times M_T}$ with the LTE system. Let $\mbf x_{\text{Radar}}(t)$ and $\mbf x_j^{\text{UE}}(t)$ be the signals transmitted from the MIMO radar and the $j^{\text{th}}$ UE in the $i^{\text{th}}$ cell, respectively. The received signal at the $i^{\text{th}}$ BS receiver can be written as
\begin{align*}
\mbf y_i(t) &= \mbf H_i^{N_R^{\text{BS}} \times M_T} \mbf x_{\text{Radar}}(t) + \sum_j \mbf H_j^{N_R^{\text{BS}} \times N_T^{\text{UE}}} \mbf x_{j}^{\text{UE}}(t) + \mbf w(t) \\
&\quad \, \, \text{for} \, \, 1 \leq i \leq N_{\text{BS}} \, \, \text{and} \, \, 1 \leq j \leq K_i^{\text{UE}}
\end{align*}
where 
$\mbf w(t)$ is the additive white Gaussian noise. In order to avoid interference to the $i^{\text{th}}$ LTE BS, the MIMO radar maps $\mbf x_{\text{Radar}}(t)$ onto the null-space of $\mbf H_i^{N_R^{\text{BS}} \times M_T}$. Figure \ref{fig:System_Model_Arch} shows a spectrum sharing scenario between a maritime MIMO radar and a LTE cellular system where the MIMO radar is sharing $N_{\text{BS}}$ interference channels $\mbf H_i^{N_R^{\text{BS}} \times M_T}$ with the LTE system.
\section{Radar-LTE Spectrum Sharing Approach}\label{sec:radar_main}
In this section, we describe colocated MIMO radars and then present the maximum likelihood (ML) estimate used as our performance metric for the MIMO radar system.
\subsection{Colocated MIMO Radar}\label{sec:radar}
The MIMO radar we consider in this paper is a colocated MIMO radar with $M_T$ transmit antennas and $M_R$ receive antennas. Let $\mbf x_{\text{Radar}}(t)$ be the signal transmitted from the MIMO radar, defined as
\begin{equation*}
\mbf x_{\text{Radar}}(t)= \begin{bmatrix} x_1(t)e^{j \omega_c t} &x_2(t)e^{j \omega_c t} &\cdots &x_{M_T}e^{j \omega_c t}(t) \end{bmatrix}^T
\end{equation*}
where 
$\omega_c$ is the carrier angular frequency, $x_k(t)$ is the baseband signal from the $k^{\text{th}}$ transmit element
%
and $t \in [0, T_o]$ with $T_o$ being the observation time. The radar transmit steering vector is defined as
\begin{equation*}
\mbf a_T(\theta) \triangleq \begin{bmatrix} e^{-j \omega_c \tau_{T_1}(\theta)} &e^{-j \omega_c \tau_{T_2}(\theta)} &\cdots &e^{-j \omega_c \tau_{T_{M_T}}(\theta)} \end{bmatrix}^T
\label{eq:at}
\end{equation*}
the radar receive steering vector is defined as
\begin{equation*}
\mbf a_R(\theta) \triangleq \begin{bmatrix} e^{-j \omega_c \tau_{R_1}(\theta)} &e^{-j \omega_c \tau_{R_2}(\theta)} &\cdots &e^{-j \omega_c \tau_{R_{M_R}}(\theta)} \end{bmatrix}^T
\label{eq:ar}
\end{equation*}
and the transmit-receive steering matrix is defined as
\begin{equation*}
\mbf A (\theta) \triangleq \mbf a_R(\theta) \mbf a_T^T(\theta).
\end{equation*}
Then, the signal received from a single point target at an angle $\theta$ is given by
\begin{equation*}
\mbf y_{\text{Radar}}(t) = \alpha \, e^{-j\omega_D t} \, \mbf A(\theta) \,  \mbf x_{\text{Radar}}(t-\tau(t)) 
\end{equation*}
where $\tau(t)=\tau_r=\tau_{T_k}(t) + \tau_{R_l}(t)$ is the sum of propagation delays between the target and the $k^{\text{th}}$ transmit element and between the target and the $l^{\text{th}}$ receive element, respectively; and $\alpha$ represents the complex path loss including the propagation loss and the coefficient of reflection.
\subsection{Maximum Likelihood}\label{sec:ML}
We choose maximum likelihood estimate of the three target parameters which are the target's angle of arrival, the propagation delay time and the Doppler angular frequency as our performance metric for the MIMO radar system. We are interested in studying the degradation in the estimate of each of these parameters due to null space projection of the radar waveform. The ML for the case of no interference and a single target can be written as in \cite{LS08},
\begin{equation}
\label{eq:ML}
(\hat{\theta},\hat{\tau}_r, \hat{\omega}_D)_{\text{ML}} = \operatorname*{arg\,max}_{\theta,\tau_r,\omega_D} \frac{\left| \mathbf{a}_R^H(\theta) \mbf E(\tau_r,\omega_D)\mathbf{a}_T^*(\theta)\right|^2}{M_R \mathbf{a}_T^H(\theta) \mathbf{R}_{\mbf x_{\text{Radar}}}^T \mathbf{a}_T(\theta)}
\end{equation}
where
\begin{align*}
\mbf R_{\mbf x_{\text{Radar}}} &= \int_{T_0} \mbf x_{\text{Radar}}(t) \, \mbf x_{\text{Radar}}^H(t) \, dt \\
\mbf E(\tau_r,\omega_D) &= \int_{T_0} \! \mathbf{y}_{\text{Radar}}(t) \, \mbf x_{\text{Radar}}^H(t-\tau_r) \, e^{j\omega_D t} \, dt
\end{align*}
$\tau_r$ is the propagation delay that refers to the two way propagation delay between the target and the reference point and $\omega_D$ is the Doppler angular frequency.

For orthogonal signal transmission, equation \eqref{eq:ML} becomes
\begin{equation}
\label{eq:ML_orthogonal}
(\hat{\theta},\hat{\tau}_r, \hat{\omega}_D)_{\text{ML}} = \operatorname*{arg\,max}_{\theta,\tau_r,\omega_D} \frac{\left| \mathbf{a}_R^H(\theta) \mbf E(\tau_r,\omega_D)\mathbf{a}_T^*(\theta)\right|^2}{M_R M_T} \cdot
\end{equation}
\section{Algorithm}\label{sec:Algorithm}
In this section, we present a channel-selection algorithm to select the best interference channel on which radar signals are projected. We also present NSP algorithm that performs the null space computation.
\subsection{Channel-Selection Algorithm}\label{sec:ICS}
\label{sec:channel}
Our channel-selection algorithm, shown in Algorithm \eqref{alg:M2C_Radar}, selects the best interference channel onto which radar signals are projected. 
Based on our system model, we assume that there exist $N_{\text{BS}}$ interference channels $\mbf H_i, i=1,2, \ldots, N_{\text{BS}}$ between the MIMO radar and the LTE system. Our goal is to select the best interference channel defined as
\begin{align*}
i_{\text{min}} &\triangleq \argmin_{1 \leq i \leq N_{\text{BS}}} \Arrowvert \mbf x_{\text{Radar}}-{\mathbf{P}}_{\mathbf{V}_i} \mbf x_{\text{Radar}} \Arrowvert \\
\mbf H_{\text{Best}} &\triangleq \mbf H_{i_{\text{min}}}
\end{align*}
we also seek to avoid the worst interference channel defined as
\begin{align*}
i_{\text{max}} &\triangleq \argmax_{1 \leq i \leq N_{\text{BS}}} \Arrowvert \mbf x_{\text{Radar}}-{\mathbf{P}}_{\mathbf{V}_i} \mbf x_{\text{Radar}} \Arrowvert \\
\mbf H_{\text{Worst}} &\triangleq \mbf H_{i_{\text{max}}}
\end{align*}
where $(\mbf x_{\text{Radar}}-{\mathbf{P}}_{\mathbf{V}_i} \mbf x_{\text{Radar}})$ is the difference between the original radar waveform $\mbf x_{\text{Radar}}$ and the radar waveform projected onto the null space of $\mbf H_i$
and the Euclidean norm of $(\mbf x_{\text{Radar}}-{\mathbf{P}}_{\mathbf{V}_i} \mbf x_{\text{Radar}})$ is defined as
\begin{align*}
\Arrowvert \mbf x_{\text{Radar}}-{\mathbf{P}}_{\mathbf{V}_i} &\mbf x_{\text{Radar}} \Arrowvert= \\
& \sqrt {(\mbf x_{\text{Radar}}-{\mathbf{P}}_{\mathbf{V}_i} \mbf x_{\text{Radar}})^H (\mbf x_{\text{Radar}}-{\mathbf{P}}_{\mathbf{V}_i} \mbf x_{\text{Radar}})}.
\end{align*}


We use the blind null space learning algorithm introduced in \cite{Blind_Null} to estimate the channel state information (CSI) of the $N_{\text{BS}}$ interference channels at the MIMO radar. The projection matrix $\mathbf{P}_{\mathbf{V}_i}$ of each of the $N_{\text{BS}}$ interference channels is then found using Algorithm \eqref{alg:NSP}. Once Algorithm \eqref{alg:M2C_Radar} receives the projection matrices of the interference channels, it selects the best interference channel $\breve{\mbf H}$ and sends it to Algorithm \eqref{alg:NSP} for NSP of radar signals. Selecting the best interference channel using our channel-selection algorithm (i.e. Algorithm \eqref{alg:M2C_Radar}) guarantees minimum degradation in the performance of the radar while maintaining no interference to the LTE BS.

\begin{algorithm}
\caption{Channel-Selection Algorithm}\label{alg:M2C_Radar}
\begin{algorithmic}
\LOOP
	\FOR{$i=1:N_{\text{BS}}$}
		\STATE{Estimate CSI of $\mbf H_i$.}
		\STATE{Send $\mbf H_i$ to Algorithm \eqref{alg:NSP} for null space computation.}
		\STATE{Receive projection matrix $\mathbf{P}_{\mathbf{V}_i}$ from  Algorithm \eqref{alg:NSP}.}
	\ENDFOR
	\STATE{Find {$i_{\text{min}} = \argmin_{1 \leq i \leq N_{\text{BS}}} \Arrowvert \mbf x_{\text{Radar}}-{\mathbf{P}}_{\mathbf{V}_i} \mbf x_{\text{Radar}} \Arrowvert$.}}

\STATE{Set $\breve{\mbf H} = \mbf H_{i_{\text{min}}}$ as the best interference channel.}
\STATE{Set $\mathbf{P}_{\breve{\mathbf{V}}} =  \mathbf{P}_{\mathbf{V}_{i_{\text{min}}}}$.}
\STATE{Send $\mathbf{P}_{\breve{\mathbf{V}}}$ to Algorithm \eqref{alg:NSP} to get NSP radar waveform.}
\ENDLOOP
\end{algorithmic}
\end{algorithm}
\subsection{Null-Space Projection (NSP) Algorithm}\label{sec:NSP}
\label{sec:nsp}
In this section, we present our proposed null-space projection algorithm. We also explain the projection of radar signals onto null space of the best interference channel selected using Algorithm \eqref{alg:M2C_Radar}. The CSI of each of the $N_{\text{BS}}$ interference channels is first estimated using a blind null space learning algorithm \cite{Blind_Null}. Algorithm \eqref{alg:NSP} gets the CSI estimates of the interference channels from Algorithm \eqref{alg:M2C_Radar} and finds the null space of each $\mbf {H}_i^{N_R^{\text{BS}} \times M_T}$. This is performed using the singular value decomposition (SVD) theorem as shown in our NSP algorithm (Algorithm \eqref{alg:NSP}). The SVD for the complex $i^{\text{th}}$ interference channel is given by
\begin{align*}
\mathbf{H}_i^{N_R^{\text{BS}} \times M_T} = \mathbf{U}_i \mathbf{\Sigma}_i^{N_R^{\text{BS}} \times M_T}\mathbf{V}_i^{H}
\end{align*}
and $\mathbf{\Sigma}_i^{N_R^{\text{BS}} \times M_T}$ is given by
\begin{align*}
{\mathbf{\Sigma}_i}^{N_R^{\text{BS}} \times M_T} = \diag(\sigma_1,...&,\sigma_k,...,\sigma_{l})\in \mathbb{R}^{N_R^{\text{BS}} \times M_T}\\ 
\textrm{s.t.}\;\;\; l&=\min \{N_R^{\text{BS}}, M_T\}
\end{align*}
where $\mbf U_i$ is the complex unitary matrix, $\bsym \Sigma_i$ is the matrix of singular values, $\sigma_1 \textgreater\sigma_2\textgreater...\textgreater\sigma_k\textgreater\sigma_{k+1}=...=\sigma_l=0$ and $\mbf V_i^H$ is the complex unitary matrix.
Once the null space of all interference channels is determined, ${\mathbf{\Sigma}_i'}^{M_T \times M_T}$ is then calculated as follows
\begin{align*}
{\mathbf{\Sigma}_i'}^{M_T \times M_T} = \diag(\sigma_1',...,\sigma_{M_T}')\in \mathbb{R}^{M_T\times M_T}\\
\textrm{s.t.}\;\;\;\;\;  \sigma_i' =  \left\{
    \begin{array}{ll}
      0 , \;\; i\leq k \\
      1 , \;\; i> k.
    \end{array}
  \right.
\end{align*}
%

Algorithm \eqref{alg:NSP} uses ${\mathbf{\Sigma}_i'}^{M_T \times M_T}$ for the formation of the projection matrix ${\mathbf{P}}_{\mathbf{V}_i}$ that is given by
\begin{eqnarray*}
\mathbf{P}_{{\mathbf{V}_i}}={\mathbf{V}_i} {\mathbf{\Sigma}_i'}^{M_T \times M_T} \mathbf{V}_i^{H} \label{Pv}
\end{eqnarray*}
where $\mathbf{P}_{{\mathbf{V}_i}}$ satisfies the following properties:
\begin{itemize}
\item $\mathbf{H}_i \mathbf{P}_{{\mathbf{V}_i}}=0$.
\item ${\mathbf{P}_{{\mathbf{V}_i}}}^2=\mathbf{P}_{{\mathbf{V}_i}}$.
\end{itemize}

Algorithm \eqref{alg:M2C_Radar} receives the projection matrices $\mathbf{P}_{{\mathbf{V}_i}}$ and uses them to determine the best interference channel $\breve{\mbf H}$ and its corresponding $\mathbf{P}_{\breve{\mathbf{V}}}$, the one with the minimum $\Arrowvert \mbf x_{\text{Radar}}-{\mathbf{P}}_{\mathbf{V}_i} \mbf x_{\text{Radar}} \Arrowvert$, which according to our Algorithm \eqref{alg:M2C_Radar} is given by
\begin{align*}
i_{\text{min}} &= \argmin_{1 \leq i \leq N_{\text{BS}}} \Arrowvert \mbf x_{\text{Radar}}-{\mathbf{P}}_{\mathbf{V}_i} \mbf x_{\text{Radar}} \Arrowvert \\
\breve{\mbf H} &= \mbf H_{i_{\text{min}}}
\\
\mathbf{P}_{\breve{\mathbf{V}}} &=  \mathbf{P}_{\mathbf{V}_{i_{\text{min}}}}.
\end{align*}

Algorithm \eqref{alg:M2C_Radar} sends $\mathbf{P}_{\breve{\mathbf{V}}}$ to Algorithm \eqref{alg:NSP} where it is used for the projection of the radar waveform. The radar waveform projected onto the null space of $\breve{\mbf H}$ can be written as
\begin{equation}
\label{eq:news}
{\breve {\mbf x}_{\text{Radar}}} = \mathbf{P}_{\breve{\mathbf{V}}} \mbf x_{\text{Radar}}.
\end{equation}

The radar projected signal given by \eqref{eq:news} can be substituted in equation \eqref{eq:ML} to get the ML estimates of the target's angle arrival, propagation delay and Doppler angular frequency for the NSP radar waveform.

\begin{algorithm}
\caption{Null-Space Projection (NSP) Algorithm}\label{alg:NSP}
\begin{algorithmic}
\IF {$\mbf H_i$ received from Algorithm \eqref{alg:M2C_Radar}}
	 \STATE{Perform SVD on $\mbf H_i$ (i.e. $\mbf H_i=\mbf U_i \bsym \Sigma_i \mbf V_i^H$).}
\STATE{Find projection matrix $\mathbf{P}_{{\mathbf{V}_i}}={\mathbf{V}_i} {\mathbf{\Sigma}_i'}^{M_T \times M_T} \mathbf{V}_i^{H}$.}
\STATE{Send projection matrix $\mathbf{P}_{{\mathbf{V}_i}}$ to Algorithm \eqref{alg:M2C_Radar}.}
\ENDIF
%
%
%
%
\IF {$\mathbf{P}_{\breve{\mathbf{V}}}$ received from Algorithm \eqref{alg:M2C_Radar}}
%
\STATE{Get NSP radar signal via ${\breve {\mbf x}_{\text{Radar}}} = \mathbf{P}_{\breve{\mathbf{V}}} \mbf x_{\text{Radar}}$.}	
\ENDIF
\end{algorithmic}
\end{algorithm}
\section{Performance Analysis of the Radar's Detectable Parameters}\label{sec:Performance}
In this section, we provide mathematical analysis to study the effect of the projected radar signals on the estimate of the radar's target detectable parameters. We compare the radar's detectable values given by the ML estimate in equation \eqref{eq:ML} for the case of original radar waveform with the radar's detectable values for the case of the NSP radar waveform. We are interested in three radar's detectable parameters that are the target's angle of arrival, the propagation delay and the Doppler angular frequency.

First we investigate the effect of projecting the radar waveform on the estimated angle of arrival. Assuming that $\tau_r$ and $\omega_D$ are known to the radar, the estimated angle of arrival $\hat{\theta}$ for the non-projected radar waveform case is given by
\begin{equation}
\label{eq:ML_theta_noNull}
\hat{\theta}_{\text{ML}} \triangleq \theta_{\text{opt}} = \operatorname*{arg\,max}_{\theta_{\text{opt}}} \frac{\left| \mathbf{a}_R^H(\theta_{\text{opt}}) \mbf E(\tau_r,\omega_D)\mathbf{a}_T^*(\theta_{\text{opt}})\right|^2}{M_R \mathbf{a}_T^H(\theta_{\text{opt}}) \mathbf{R}_{{\mbf x}_{\text{Radar}}}^T \mathbf{a}_T(\theta_{\text{opt}})}
\end{equation}
whereas in the case of the projected radar waveform the ML estimate for the angle of arrival can be written as
\begin{equation}
\label{eq:ML_theta_Null}
\begin{aligned}
\hat{\theta}_{\text{ML}^{\text{NSP}}} \triangleq \theta_{\text{opt}}^{\text{NSP}}
&= \operatorname*{arg\,max}_{\theta_{\text{opt}}^{\text{NSP}}} \frac{\left| \mathbf{a}_R^H(\theta_{\text{opt}}^{\text{NSP}}) \mbf E^{\text{NSP}}(\tau_r,\omega_D) \mathbf{a}_T^*(\theta_{\text{opt}}^{\text{NSP}})\right|^2}{M_R \mathbf{a}_T^H(\theta_{\text{opt}}^{\text{NSP}}) {\mathbf{R}_{{\mbf x}_{\text{Radar}}}^{T^{\text{NSP}}}} \mathbf{a}_T(\theta_{\text{opt}}^{\text{NSP}})}\\
&= \operatorname*{arg\,max}_{\theta_{\text{opt}}^{\text{NSP}}} \frac{\left| \mathbf{a}_R^H(\theta_{\text{opt}}^{\text{NSP}}) \mathbf{P}_{\breve{\mathbf{V}}} \mbf E(\tau_r,\omega_D) \mathbf{P}_{\breve{\mathbf{V}}}^H \mathbf{a}_T^*(\theta_{\text{opt}}^{\text{NSP}})\right|^2}{M_R \mathbf{a}_T^H(\theta_{\text{opt}}^{\text{NSP}}) \mathbf{P}_{\breve{\mathbf{V}}} \mathbf{R}_{{\mbf x}_{\text{Radar}}}^T \mathbf{P}_{\breve{\mathbf{V}}}^H \mathbf{a}_T(\theta_{\text{opt}}^{\text{NSP}})}\\
&= \operatorname*{arg\,max}_{\theta_{\text{opt}}^{\text{NSP}}} \frac{\left| \mathbf{a}_R^H(\theta_{\text{eff}}^{\text{NSP}}) \mbf E(\tau_r,\omega_D) \mathbf{a}_T^*(\theta_{\text{eff}}^{\text{NSP}})\right|^2}{M_R \mathbf{a}_T^H(\theta_{\text{eff}}^{\text{NSP}}) \mathbf{R}_{{\mbf x}_{\text{Radar}}}^T \mathbf{a}_T(\theta_{\text{eff}}^{\text{NSP}})}
\end{aligned}
\end{equation}
where $\mathbf{a}_R^H(\theta_{\text{eff}}^{\text{NSP}})=\mathbf{a}_R^H(\theta_{\text{opt}}^{\text{NSP}})\mathbf{P}_{\breve{\mathbf{V}}}$, $\mathbf{a}_T^*(\theta_{\text{eff}}^{\text{NSP}})=\mathbf{P}_{\breve{\mathbf{V}}}^H \mathbf{a}_T^*(\theta_{\text{opt}}^{\text{NSP}})$, $\mathbf{a}_T^H(\theta_{\text{eff}}^{\text{NSP}})=\mathbf{a}_T^H(\theta_{\text{opt}}^{\text{NSP}}) \mathbf{P}_{\breve{\mathbf{V}}}$ and $\mathbf{a}_T(\theta_{\text{eff}}^{\text{NSP}})=\mathbf{P}_{\breve{\mathbf{V}}}^H \mathbf{a}_T(\theta_{\text{opt}}^{\text{NSP}})$. Also $\mbf E^{\text{NSP}}(\tau_r,\omega_D)=\mathbf{P}_{\breve{\mathbf{V}}} \mbf E(\tau_r,\omega_D) \mathbf{P}_{\breve{\mathbf{V}}}^H$ and $\mathbf{R}_{{\mbf x}_{\text{Radar}}}^{T^{\text{NSP}}}=\mathbf{P}_{\breve{\mathbf{V}}} \mathbf{R}_{{\mbf x}_{\text{Radar}}}^T \mathbf{P}_{\breve{\mathbf{V}}}^H$. Equation \eqref{eq:ML_theta_Null} shows that the estimated angle of arrival $\theta_{\text{opt}}^{\text{NSP}}$ in the case of the NSP radar waveform is different from the estimated angle of arrival $\theta_{\text{opt}}$ given by \eqref{eq:ML_theta_noNull} in the case of the original radar waveform. Section \ref{section:sim} shows that by choosing $\mbf H_{\text{Best}}$ to project using Algorithm \eqref{alg:M2C_Radar} and \eqref{alg:NSP} we can achieve almost similar ML results in estimating the angle of arrival for the original radar waveform and the NSP waveform.

The effect of the projected radar waveform in estimating the propagation delay $\tau_r$ is also investigated. Assuming that $\theta$ and $\omega_D$ are known to the radar, the estimated propagation delay $\hat{\tau}_r$ for the non-projected radar waveform case is given by
\begin{equation}
\label{eq:ML_delay_noNull}
(\hat{\tau}_r)_{\text{ML}} \triangleq {\tau_{r,\text{opt}}} = \operatorname*{arg\,max}_{{\tau_{r,\text{opt}}}} \frac{\left| \mathbf{a}_R^H(\theta) \mbf E({\tau_{r,\text{opt}}},\omega_D)\mathbf{a}_T^*(\theta)\right|^2}{M_R \mathbf{a}_T^H(\theta) \mathbf{R}_{{\mbf x}_{\text{Radar}}}^T \mathbf{a}_T(\theta)}
\end{equation}
where $\mbf E(\tau_{r,\text{opt}},\omega_D) = \int_{T_0} \! \mathbf{y}_{\text{Radar}}(t) \, \mbf x_{\text{Radar}}^H(t-\tau_{r,\text{opt}}) \, e^{j\omega_D t} \, dt$ is equivalent to $\int_{T_0} \! \alpha \, e^{-j\omega_D t} \, \mbf A(\theta) \,  \mbf x_{\text{Radar}}(t-\tau) \, \mbf x_{\text{Radar}}^H(t-\tau_{r,\text{opt}}) \, e^{j\omega_D t} \, dt$. Since the radar waveforms are orthogonal waveforms, the objective function in equation \eqref{eq:ML_delay_noNull} can be maximized when $\mbf E(\tau_{r,\text{opt}},\omega_D)$ is maximized and can be achieved when  $\tau_{r,\text{opt}}=\tau$. In the case of the projected radar waveform the ML estimate for the propagation delay can be written as
\begin{equation}
\label{eq:ML_delay_Null}
(\hat{\tau}_r)_{\text{ML}^{\text{NSP}}} \triangleq {\tau_{r,\text{opt}}^{\text{NSP}}} = \operatorname*{arg\,max}_{\tau_{r,\text{opt}}^{\text{NSP}}} \frac{\left| \mathbf{a}_R^H(\theta) \mathbf{P}_{\breve{\mathbf{V}}} \mbf E({\tau_{r,\text{opt}}^{\text{NSP}}},\omega_D) \mathbf{P}_{\breve{\mathbf{V}}}^H \mathbf{a}_T^*(\theta)\right|^2}{M_R \mathbf{a}_T^H(\theta) \mathbf{P}_{\breve{\mathbf{V}}} \mathbf{R}_{{\mbf x}_{\text{Radar}}}^T \mathbf{P}_{\breve{\mathbf{V}}}^H \mathbf{a}_T(\theta)} \cdot
\end{equation}

Again the objective function in \eqref{eq:ML_delay_Null} can be maximized when $\mbf E(\tau_{r,\text{opt}}^{\text{NSP}},\omega_D)$ is maximized and can be achieved when  $\tau_{r,\text{opt}}^{\text{NSP}}=\tau$. Therefore, projecting the radar waveforms has no effect on the estimated propagation delay and equation \eqref{eq:ML_delay_Null} gives an estimate for the propagation delay similar to the one given by equation \eqref{eq:ML_delay_noNull}.

Furthermore, the effect of the projected radar waveform in estimating the Doppler angular frequency $\omega_D$ is also investigated. Assuming that $\theta$ and $\tau_r$ are known to the radar, the Doppler angular frequency $\hat{\omega}_D$ for the non-projected radar waveform case is given as
\begin{equation}
\label{eq:ML_doppler_noNull}
(\hat{\omega}_D)_{\text{ML}} \triangleq {\omega_{D,\text{opt}}} = \operatorname*{arg\,max}_{{\omega_{D,\text{opt}}}} \frac{\left| \mathbf{a}_R^H(\theta) \mbf E({\tau_r},\omega_{D,\text{opt}})\mathbf{a}_T^*(\theta)\right|^2}{M_R \mathbf{a}_T^H(\theta) \mathbf{R}_{{\mbf x}_{\text{Radar}}}^T \mathbf{a}_T(\theta)}
\end{equation}
where $\mbf E(\tau_r,\omega_{D,\text{opt}}) = \int_{T_0} \! \mathbf{y}_{\text{Radar}}(t) \, \mbf x_{\text{Radar}}^H(t-\tau_r) \, e^{j\omega_{D,\text{opt}} t} \, dt$ is equivalent to $\int_{T_0} \! \alpha \, e^{-j\omega_D t} \, \mbf A(\theta) \,  \mbf x_{\text{Radar}}(t-\tau) \, \mbf x_{\text{Radar}}^H(t-\tau_r) \, e^{j\omega_{D,\text{opt}} t} \, dt$ and $\tau=\tau_r$ in this case. Since the radar waveforms are orthogonal waveforms, the objective function in equation \eqref{eq:ML_delay_noNull} can be maximized when $\mbf E(\tau_r,\omega_{D,\text{opt}})$ is maximized and can be achieved when  $\omega_{D,\text{opt}}=\omega_D$. In the case of the projected radar waveform the ML estimate for the propagation delay can be written as
\begin{equation}
\label{eq:ML_doppler_Null}
(\hat{\omega}_D)_{\text{ML}^{\text{NSP}}} \triangleq {\omega_{D,\text{opt}}^{\text{NSP}}} = \operatorname*{arg\,max}_{\omega_{D,\text{opt}}^{\text{NSP}}} \frac{\left| \mathbf{a}_R^H(\theta) \mathbf{P}_{\breve{\mathbf{V}}} \mbf E({\tau_r},\omega_{D,\text{opt}}^{\text{NSP}}) \mathbf{P}_{\breve{\mathbf{V}}}^H \mathbf{a}_T^*(\theta)\right|^2}{M_R \mathbf{a}_T^H(\theta) \mathbf{P}_{\breve{\mathbf{V}}} \mathbf{R}_{{\mbf x}_{\text{Radar}}}^T \mathbf{P}_{\breve{\mathbf{V}}}^H \mathbf{a}_T(\theta)} \cdot
\end{equation}

The objective function in \eqref{eq:ML_doppler_Null} can be maximized when $\mbf E(\tau_{r,\text{opt}}^{\text{NSP}},\omega_D)$ is maximized and can be achieved when  $\omega_{D,\text{opt}}^{\text{NSP}}=\omega_D$. Therefore, projecting the radar waveforms has no effect on the estimated Doppler angular frequency and equation \eqref{eq:ML_doppler_Null} gives an estimate for the propagation delay that is similar to the one given by equation \eqref{eq:ML_doppler_noNull}.
\section{Simulations}
\label{section:sim}
In this section, we simulate the presented MIMO radar and LTE spectrum sharing scenario and study its impact on the performance of the radar in estimating the detectable target parameters. The simulation parameters used are listed in Table \ref{tab1}. Algorithm \eqref{alg:M2C_Radar} and Algorithm \eqref{alg:NSP} were applied in MATLAB to select the best interference channel $\mbf H_{\text{Best}}$ and project the radar signal onto it. By applying Algorithm \eqref{alg:M2C_Radar} and Algorithm \eqref{alg:NSP}, we are able to minimize the degradation in the radar performance as projecting the radar waveform onto $\mbf H_{\text{Best}}$ results in a NSP waveform that is closer to the original radar waveform than the NSP waveform in the case of projecting the radar waveform onto $\mbf H_{\text{Worst}}$. We compared the radar estimated detectable target parameters, using ML estimate, in the case of the original radar waveform and the two cases of NSP radar waveform, first when choosing $\mbf H_{\text{Best}}$ to project and second when choosing $\mbf H_{\text{Worst}}$ to project.

\renewcommand{\arraystretch}{1.5}
\begin{table}
\centering
\caption{MIMO Radar System Parameters}
\begin{tabular}{lll}
  \topline
  \headcol Parameters &Notations &Values\\
  \midline
  		    Radar/LTE shared RF band &- &$3550-3650$ MHz \\
    \rowcol Radar waveform bandwidth &$B$ &10 MHz \\
	        Radar transmit antennas & $M_T$ &10 \\
	\rowcol Radar receive antennas & $M_R$ &7 \\
 			Carrier frequency &$f_c$ &3.55 GHz \\
	\rowcol Wavelength &$\lambda$ &8.5 cm \\
			Inter-element antenna spacing &$3\lambda/4$ &6.42 cm \\
	\rowcol Radial velocity & $v_r$ &2000 m/s \\
			Speed of light &$c$ & 3 $\times$ $10^8$ m/s\\
	\rowcol	Observation time &$T_0$ &1 ms\\
			Target distance from the radar & $r_0$ &5000 m \\
	\rowcol Target angle &$\theta$ & $0^{\circ}$ \\
            Doppler frequency shift &$f_d$  & $2 v_r/\lambda$  \\
	\rowcol	Doppler angular frequency &$\omega_D$  & $2 \omega_c v_r/c=2 \pi f_d$  \\
			propagation delay &$\tau_r$ & $2 r_0/c$ \\
	\rowcol	Path loss &$\alpha$  & $\alpha_{ji} e^{-j \omega_c \tau_r}$ \\
	
  \hline
\end{tabular}
\label{tab1}
\end{table}

\begin{figure}
\includegraphics[height=2in, width=3.5in]{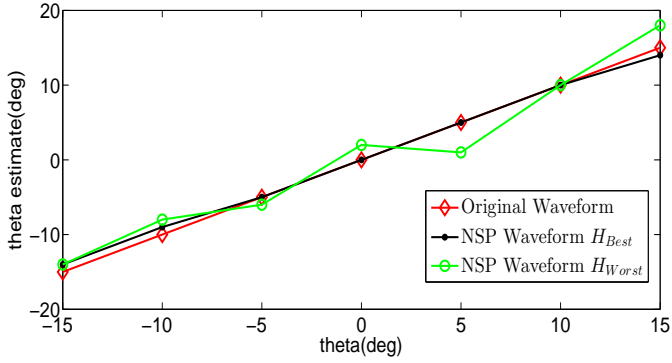}
\caption{Target direction $\hat{\theta}$ estimation using ML estimate in the case of the original radar waveform, the NSP radar waveform projected onto $\mbf H_{\text{Best}}$ and the NSP radar waveform projected onto $\mbf H_{\text{Worst}}$.}
\myfigureshrinker{\vspace{-0.06in}}
\label{fig:theta_estimation}
\end{figure}

\begin{figure}
\includegraphics[height=1.8in, width=3.5in]{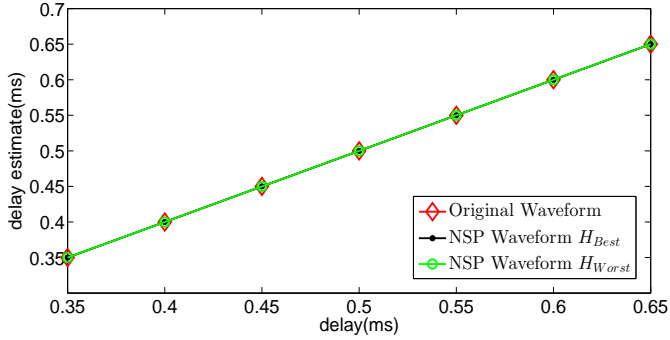}
\caption{Propagation delay $\hat{\tau}_r$ estimation using ML estimate in the case of the original radar waveform, the NSP radar waveform projected onto $\mbf H_{\text{Best}}$ and the NSP radar waveform projected onto $\mbf H_{\text{Worst}}$.}
\myfigureshrinker{\vspace{-0.06in}}
\label{fig:delay_estimation}
\end{figure}

\begin{figure}
\includegraphics[height=1.8in, width=3.5in]{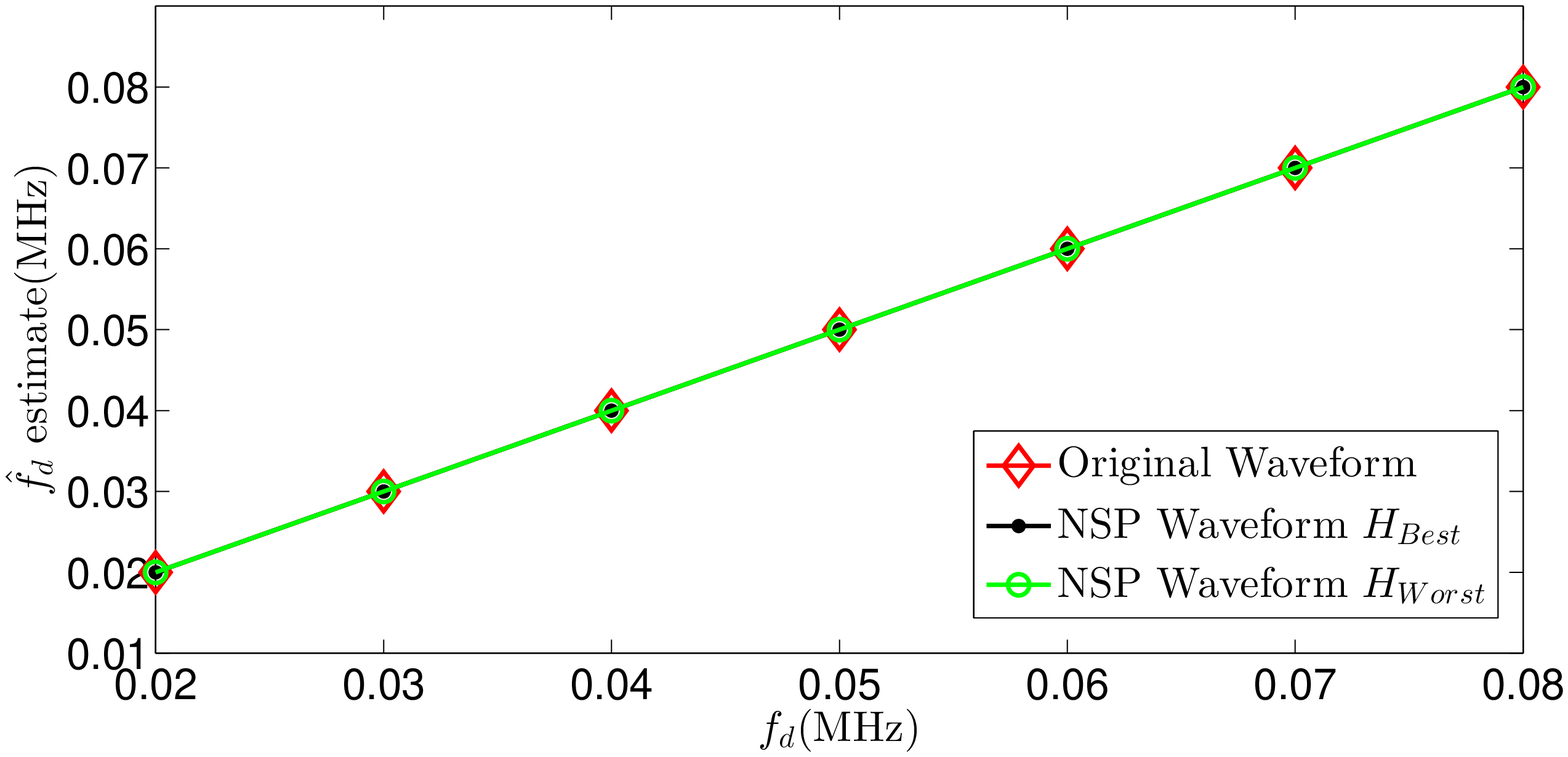}
\caption{Doppler frequency shift $\hat{f}_d$ estimation using ML estimate in the case of the original radar waveform, the NSP radar waveform projected onto $\mbf H_{\text{Best}}$ and the NSP radar waveform projected onto $\mbf H_{\text{Worst}}$.}
\myfigureshrinker{\vspace{-0.06in}}
\label{fig:doppler_estimation}
\end{figure}

In Figure \ref{fig:theta_estimation}, we compare the estimated angles when using the ML estimate described in equation \eqref{eq:ML} for the original radar waveform with the estimated angles using the ML estimate for both of the NSP radar waveform onto $\mbf H_{\text{Best}}$ and the NSP radar waveform onto $\mbf H_{\text{Worst}}$. Figure \ref{fig:theta_estimation} shows that choosing $\mbf H_{\text{Best}}$ to project can cause less degradation in estimating the angle of arrival than choosing $\mbf H_{\text{Worst}}$ to project.
In Figure \ref{fig:delay_estimation}, we compare the estimated propagation delay $\hat{\tau}_r$ when using the ML estimate for the original radar waveform with the estimated propagation delay using the ML estimate for both of the NSP radar waveform onto $\mbf H_{\text{Best}}$ and the NSP radar waveform onto $\mbf H_{\text{Worst}}$. Figure \ref{fig:delay_estimation} shows that projecting the radar waveform does not affect the estimated propagation delay and results in an estimated value that is similar to the one obtained when using the original radar waveform. Similarly, Figure \ref{fig:doppler_estimation} shows that the estimated Doppler frequency shift $\hat{f}_d$ when using the projected radar waveform, i.e. for the case of NSP radar waveform onto $\mbf H_{\text{Best}}$ and the case of NSP radar waveform onto $\mbf H_{\text{Worst}}$ is similar to the estimated $\hat{f}_d$ when using the original radar waveform.

\section{Conclusion}
\label{section:conclude}

In this paper, we presented a spectrum sharing scenario between a MIMO radar and LTE cellular system with multiple BSs. We proposed a channel-selection algorithm and NSP algorithm to select the best interference channel and project the radar signal onto it. Our proposed algorithms guarantee a minimum degradation in the radar's performance by selecting the best interference channel for the NSP of the radar signal. We showed through mathematical analysis the effect of projecting the radar signal on the radar's target detectable parameters. Our analysis showed that when using the ML estimate, the estimated propagation delay and the estimated Doppler angular frequency in the case of the original radar waveform are similar to the estimated values in the case of the projected radar waveform whereas the estimated angle of arrival is affected when projecting the radar waveform. We showed through our simulation results that the estimated propagation delay and Doppler frequency shift are not affected when using the NSP radar waveforms and that the loss in the radar performance in detecting the angle of arrival is minimal when our proposed spectrum sharing algorithm is used to select the best channel onto which radar signals are projected.
\bibliographystyle{ieeetr}
\bibliography{pubs}
\end{document}